\documentstyle[aps,prl,preprint,floats,epsfig]{revtex}

\def\EPJ{{Eur. Phys. Jour.} C\,}

\def\NIMA{{Nucl. Instrum. Methods Phys. Res.} A\,}
\def\NPB{{Nucl. Phys. B\,}}

\def\PRD{{Phys. Rev.} D}
\def\PRL{{Phys. Rev. Lett.}}

\def\CLEO{CLEO Collaboration}
\def\etal{{\it et al.}}

\providecommand{\mm}{\mbox{\rm mm}}

\providecommand{\invfb}{\mbox{${\rm fb^{-1}}$}}

\providecommand{\MeV}{\mbox{\rm MeV}}
\providecommand{\MeVc}{\mbox{${\rm MeV}/c$}}
\providecommand{\MeVcsq}{\mbox{${\rm MeV}/c^2$}}
\providecommand{\GeV}{\mbox{\rm GeV}}
\providecommand{\GeVc}{\mbox{${\rm GeV}/c$}}
\providecommand{\GeVcsq}{\mbox{${\rm GeV}/c^2$}}

\providecommand{\T}{\mbox{\rm T}}
\providecommand{\CL}{\mbox{\rm C.L.}}

\providecommand{\ra}{\mbox{$\rightarrow$}}

\providecommand{\epl}{\mbox{$e^+$}}
\providecommand{\emi}{\mbox{$e^-$}}
\providecommand{\eplemi}{\mbox{\epl \emi}}

\providecommand{\pipl}{\mbox{$\pi^+$}}
\providecommand{\pimi}{\mbox{$\pi^-$}}
\providecommand{\pizero}{\mbox{$\pi^0$}}
\providecommand{\piplpimi}{\mbox{\pipl \pimi}}

\providecommand{\qbar}{\mbox{$\overline{q}$}}

\providecommand{\cbar}{\mbox{$\overline{c}$}}

\providecommand{\bbar}{\mbox{$\overline{b}$}}

\providecommand{\Kzero}{\mbox{$K^0$}}

\providecommand{\KS}{\mbox{$K^0_S$}}

\providecommand{\bbar}{\mbox{\bbar}} 
\providecommand{\cbar}{\mbox{\cbar}}
\providecommand{\qbar}{\mbox{\qbar}}

\providecommand{\degrees}{\mbox{$^\circ$}}

\providecommand{\Vub}{\mbox{$V_{ub}$}}

\providecommand{\etac}{\mbox{$\etac(1S)$}}

\providecommand{\Dzero}{\mbox{$D^0$}}

\providecommand{\Dpl}{\mbox{$D^+$}}

\providecommand{\Dstarpl}{\mbox{$D^{*+}$}}

\providecommand{\Bbar}{\mbox{$\overline{B}$}}
\providecommand{\BBbar}{\mbox{$B \overline{B}$}}                                                   

\providecommand{\Bpl}{\mbox{$B^+$}}
\providecommand{\Bmi}{\mbox{$B^-$}}

\begin{document}

\preprint{\tighten\vbox{\hbox{\hfil CLNS 01/1720}
                        \hbox{\hfil CLEO 01-4}
}}

\title{Search for the Decay $\Bpl \ra \Dstarpl \KS$}

\author{CLEO Collaboration}
\date{February 14, 2001}

\maketitle
\tighten

\begin{abstract} 
We report on a search for the decay $\Bpl \ra \Dstarpl \KS$ and its
charge conjugate with the CLEO detector at the Cornell Electron Storage
Ring (CESR).
No candidates are found in $9.10\, \invfb$ of data.
The background is estimated to be $0.29 \pm 0.05$ leading to an upper limit 
${\cal B} (\Bpl \ra \Dstarpl \Kzero) = 9.5 \times 10^{-5}$ $(90\%\, \CL)$. 
\end{abstract}
\newpage

{
\renewcommand{\thefootnote}{\fnsymbol{footnote}}

\begin{center}
A.~Gritsan,$^{1}$
J.~P.~Alexander,$^{2}$ R.~Baker,$^{2}$ C.~Bebek,$^{2}$
B.~E.~Berger,$^{2}$ K.~Berkelman,$^{2}$ F.~Blanc,$^{2}$
V.~Boisvert,$^{2}$ D.~G.~Cassel,$^{2}$ P.~S.~Drell,$^{2}$
% ...
\end{center}

 
\small
\begin{center}
$^{1}${University of Colorado, Boulder, Colorado 80309-0390}\\
$^{2}${Cornell University, Ithaca, New York 14853}\\
% ...
\end{center}

\setcounter{footnote}{0}
}
\newpage


\section{Introduction}

The decay $\Bpl \ra \Dstarpl \Kzero$ (throughout this report charge
conjugate decays are implied) is expected to proceed through an
annihilation diagram with a $W^+$ in the $s$-channel
with a rate proportional to $|\Vub|^2$.
Although no calculation exists for the rate of $\Bpl \ra \Dstarpl \KS$, 
the branching fraction of the related decay $\Bpl \ra \Dpl \Kzero$
has been estimated~\cite{ref:Xing} to be in the range $0.8 \times 10^{-8}$ 
to $3 \times 10^{-6}$.
The main uncertainty in the calculation arises from the unknown 
contribution from re-scattering of the final state particles.
In the reaction $\eplemi \ra \Upsilon(4S) \ra \Bpl \Bmi$ the $B$
mesons are produced nearly at rest in the laboratory.
Therefore the $\Dstarpl$ and the $\KS$ daughters have large 
momenta of order of $2.2\, \GeVc$ essentially in opposite directions.
Background events from multi-body charm decay and light quark fragmentation 
generally do not reconstruct to back-to-back $\Dstarpl$ and $\KS$ pairs
with momenta near $2.2\, \GeVc$.
Background rejection is further helped by the excellent resolution of 
the $\Dstarpl {\text -}\, \Dzero$ mass difference and by the reconstruction
of the $\Kzero$ as a $\KS$ with excellent mass resolution and a significant
decay length.

\section{Detector and Datasets}

The CLEO detector\,\cite{ref:CLEOdet} is a general purpose detector 
that provides charged particle tracking, 
precision electromagnetic calorimetry,
charged particle identification, and muon detection.
Charged particle detection over $95\%$ of the solid angle is achieved 
by tracking devices in two different configurations, 
situated in a magnetic field of $1.5\, \T$.
In the first configuration (CLEO II), tracking is provided by three 
concentric wire chambers while in the second configuration (CLEO II.V)
the innermost wire chamber is replaced by a precision
three-layer silicon vertex detector\cite{ref:SVX}.
The momentum resolution is 0.5\% at $p = 1\, \GeVc$.
The drift chambers are surrounded by a time of flight (TOF) system.
Energy loss ($dE/dx$) in the outer drift chamber 
and the TOF system provide pion-kaon separation.
A CsI based electromagnetic calorimeter consisting of a barrel and two
endcaps (boundaries at $45\degrees$ with respect to the beams) 
has an energy resolution of 4\% for $100\, \MeV$ electromagnetic 
showers and provides $\pizero$ detection.
A superconducting coil and muon detectors surround the calorimeter.
Redundant triggers provide efficient registration of mutiparticle
final states.

The Cornell Electron Store Ring (CESR) operates at a center-of-mass energy 
of approximately $10.6\, \GeV$.
The results in this report are
based upon $9.10\, \invfb$ of integrated luminosity produced at $\eplemi$
center-of-mass energy on the $\Upsilon(4S)$.  
An additional $4.29\, \invfb$ produced $60\, \MeV$ below the $B \Bbar$ 
threshold provides an estimate of the background due to 
$\eplemi \ra q \qbar$, where $q = u, d, s, c$.  
Hereafter, we refer to these two data samples as 'on-4$S$' 
and 'off-4$S$' data, respectively.
The number of $B \Bbar$ pairs is $(9.63 \pm 0.19) \times 10^6$.

The Monte Carlo simulation of the CLEO detector is based upon 
GEANT\cite{ref:GEANT}.
Simulated events are processed in the same manner as the data.

\section{Event Reconstruction}

Charged pion and kaon candidates are selected from tracks that are well 
reconstructed, consistent with originating from the $\eplemi$ interaction 
point, and not identified as a muon.
Particle identification is used to identify charged pions and kaons.
The photons used in the $\pizero$ reconstruction are required to have an 
energy greater than 30 and $50\, \MeV$ in the barrel and endcap regions, 
respectively, and to not be associated with a charged track.
The photon-photon invariant mass is required to be within 
three standard deviations of the known\cite{ref:PDG} $\pizero$ mass.
The $\pizero$ are constrained to their known mass
and their momentum is required to be greater than $100\, \MeVc$.
A mass constrained fit is applied to $\KS$ candidates that are formed 
from oppositely charged pions.
The fit is required to have a $\chi^2$ less than 10, improving the resultant
$\KS$ momentum resolution by 5\% for $\KS$ from $\Dzero$ decays while no 
significant improvement results for $\KS$ from $\Bpl$ decays.
The $\KS$ candidates are required to originate from the $\eplemi$ 
interaction point and to have a significant decay path (at least three and 
five standard deviations in the CLEO II and CLEO II.V configurations 
respectively).
The decay path is measured with typical standard deviations of 
$1.2\, \mm$ in CLEO II and $0.7\, \mm$ in CLEO II.V.

$\Dzero$ candidates are reconstructed in five decay channels: 
$K^- \pipl$, $K^- \pipl \pizero$, $K^- \pipl \pipl \pimi$,
$\KS\ \pipl \pimi$, and $\KS\ \pipl \pimi \pizero$.
The charged $\Dzero$ daughters are constrained to a common vertex and 
the momentum of the $\Dzero$ candidate is required to be larger than 
$1.1\, \GeVc$. 
These $\Dzero$ candidates are paired with charged pions to form 
$\Dstarpl$ candidates which are constrained 
to their known\cite{ref:PDG} mass, 
thus improving their momentum resolution by approximately $14\%$.
The $\Dstarpl$ candidates are required to a momentum larger than 
$1.3\, \GeVc$.

\section{Event Selection}

Background is predominantly due to non-$B\Bbar$ sources so the 
off-$4S$ data provide a good monitor of the requirements' effectiveness 
in rejecting background.
Signal event selection requirements are defined using simulated signal events 
and off-$4S$ data without reference to on-$4S$ data.
Signal event selection variables and the corresponding requirements are: 
$\chi_m^2 \le 3.5$ (defined in Eq.\, 1 below), 
$|\cos{\theta_{\rm thr}}| \le 0.9$ ($\theta_{\rm thr}$ is the angle
between the thrust axis\cite{ref:Thrust} of the $B$ candidate and the 
thrust axis of the rest of the event),  
the normalized second Fox-Wolfram moment\cite{ref:FW} $\le 0.3$, and 
$|\cos{\theta_{\rm hel}}| \ge 0.5$ ($\theta_{\rm hel}$ is the 
helicity angle defined as the angle between the $\pipl$ from $\Dstarpl$ 
decay and the $\Dstarpl$ direction in the $\Dzero$ rest frame).

$\chi_m^2$ is defined as 
\begin{eqnarray}
\chi^2_m & \equiv &
\left( \frac {m_D - m^{\rm nom}_D} 
{\sigma(m_D)} \right)^2 + 
\left( \frac {\Delta m_{D^* D} - 
\Delta m_{D^* D}^{\rm nom}} 
{\sigma(\Delta m_{D^* D^0})} \right)^2 \nonumber \\
& & + \left( \frac {m_{\pi\pi} - m^{\rm nom}_{K_S}}
{\sigma(m_{\pi\pi})} \right)^2
\end{eqnarray}
Here, $m_D$ is the invariant mass of the $\Dzero$ candidate,
$\Delta m_{D^* D}$ is the mass difference between the 
$\Dstarpl$ and the $\Dzero$ candidates,
and $m_{\pi\pi}$ is the invariant mass
of $\piplpimi$ pairs that form the $\KS$ candidate;
all are calculated from the unconstrained kinematics 
of the respective final state particles.
The same quantities labeled by "${\rm nom}$" are their 
known\cite{ref:PDG} values.
The $\sigma(m_D)$ and $\sigma(\Delta m_{D^* D})$ are the 
per $\Dzero$ decay channel resolutions of $m_D$ and 
$\Delta m_{D^* D}$ respectively
and are determined from the data.
They are approximately $6\, \MeVcsq$ and $0.53\, \MeVcsq$ respectively.
The efficiency of the four requirements is $34\, \%$ and their
background rejection factor is 82.
If an event has more than one $\Bpl \ra \Dstarpl \KS$ candidate,
the candidate with the lowest $\chi_m^2$ is chosen.

\section{Results}

\begin{figure}[htbp]
\begin{center}
\epsfxsize=8.0cm
\epsfbox{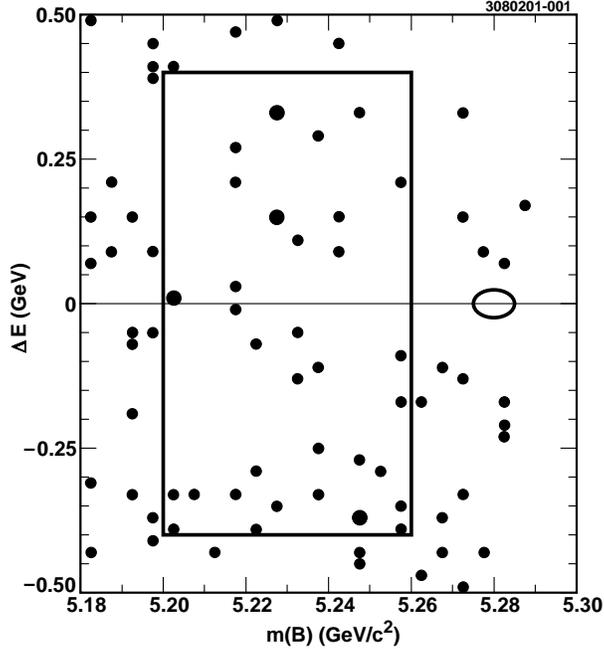}
\caption{\label{fig:dEmBon}
The $\Delta E - m(B)$ distribution for on-$4S$ events.
The signal ellipse and the rectangular background region are shown.}
\end{center}
\end{figure}
Results are presented in a two-dimensional plot of the energy difference
$\Delta E$ and the beam-constrained mass $m(B)$, 
see Fig.~\ref{fig:dEmBon}, 
with $\Delta E = E_{D^{*+}} + E_{K_S^\circ} - E_{\rm beam}$ and
$m(B) = \sqrt{E_{\rm beam}^2 - ({\bf p}_{D^*} + {\bf p}_{K_S})^2}$.
Here $E_{D^{*+}}$, $E_{K_S^\circ}$, ${\bf p}_{D^*}$, 
and ${\bf p}_{K_S}$ are the mass-constrained energy and momentum 
of the $\Dstarpl$ and the $\KS$ respectively.
The resolution in $\Delta E$ is $11.9\, \MeV$ and in $m(B)$ $2.5\,\MeVcsq$,
the latter dominated by the spread in beam energy.
The signal region is enclosed by an ellipse with semiaxes of length 
$24\, \MeV$ along $\Delta E$ and $5\, \MeVcsq$ along $m(B)$ with 
an efficiency of $82\, \%$.
There are no events in the signal region.
The background in the signal region is estimated from the number of
events that are in the rectangular region shown in Fig.~\ref{fig:dEmBon},
defined by $|\Delta E| < 0.4\, \GeV$ and $5.2 \le m(B) \le 5.26\, \GeVcsq$.
The 37 events in this region are scaled by the ratio of areas of the signal 
and background region to give a background estimate of $0.29 \pm 0.05$ events.

\begin{figure}[htbp]
\begin{center}
\epsfxsize=8.0cm
\epsfbox{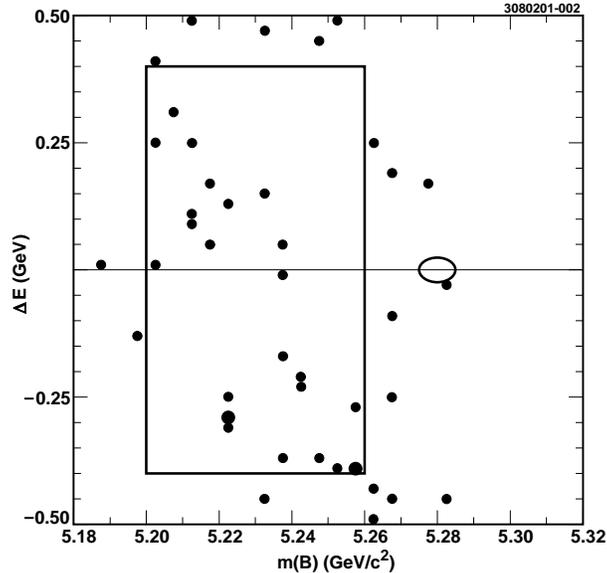}
\caption{\label{fig:dEmBoff}
The $\Delta E - m(B)$ distribution for off-$4S$ events.
The signal ellipse and the rectangular background region are shown.}
\end{center}
\end{figure}
We show in Fig.~\ref{fig:dEmBoff} the $\Delta E - m(B)$ distribution 
for off-$4S$ events.
Here the calculated value of $m(B)$ is increased by $30\, \MeVcsq$ to take 
into account the lower center-of-mass energy of the off-$4S$ data.
There are no signal events in the signal region and 25 events 
in the background region.
To compare with the 37 events found in the on-$4S$ data, 
the 25 events must be scaled by the ratio of luminosities in
the on-$4S$ and off-$4S$ data and the ratio of the
center-of-mass energy squared to obtain $50.1 \pm 10.0$.
This number is larger than but consistent with the 37 events in the background 
region in on-4$S$ data and shows that the background is predominantly 
from non-$B\Bbar$ sources.

Based upon an acceptance (including all branching fractions) 
of $2.76 \times 10^{-3}$ and the number of $B \Bbar$ pairs, 
one signal event corresponds to 
${\cal B}(\Bpl \ra \Dstarpl \Kzero) = 3.7 \times 10^{-5}$.
Here we assumed equal production of neutral and charged $B$ pairs from
$\Upsilon(4S)$ decays, consistent\cite{ref:fplmi} with experiment.
The observation of zero events in the signal ellipse with an expectation
of $0.29$ events from background corresponds~\cite{ref:Feldman} to a 
$90\%\, \CL$ upper limit on the the number of signal events of 2.15 and
a $90\%\, \CL$ upper limit on the branching fraction of 
${\cal B} (\Bpl \ra \Dstarpl \Kzero) = 8.0 \times 10^{-5}$.

Systematic uncertainties originate from track finding (1\% per track, 
5\% per soft pion from $\Dstarpl$ decay), 
$dE/dx$ (2\% per track), $\KS$ finding (3\% per $\KS$), 
$\pizero$ finding (5.5\% per $\pizero$), 
the number of produced $B\Bbar$ (2\%), 
and the five $\Dzero$ branching fractions (2.0-5.2\%\cite{ref:PDG} 
for the three dominant $\Dzero$ decay channels
$K^- \pipl$, $K^- \pipl \pizero$, $K^- \pipl \pipl \pimi$).
The weighted average systematic uncertainty is 10\%.
The 16\% statistical uncertainty in the background estimate is added 
to obtain a total uncertainty on the background estimate of 19\%. 
This leads to a $90\%\, \CL$ upper limit 
${\cal B}(\Bpl \ra \Dstarpl \Kzero) = 9.5 \times 10^{-5}$.

\section{Summary and Conclusion}

We have searched for the decay $\Bpl \ra \Dstarpl \KS$.
In a data sample with $9.10\, \invfb$ integrated luminosity,
corresponding to $9.63 \times 10^{6}$ produced $\BBbar$ pairs, 
we found zero signal events on an estimated background of $0.29 \pm 0.05$.
Including systematic uncertainties, we find an upper limit 
${\cal B} (\Bpl \ra \Dstarpl \Kzero) = 9.5 \times 10^{-5}$ $(90\%\, \CL)$.

\section{Acknowledgements}

We gratefully acknowledge the effort of the CESR staff in providing us with
excellent luminosity and running conditions.
M. Selen thanks the PFF program of the NSF and the Research Corporation, 
A.H. Mahmood thanks the Texas Advanced Research Program,
F. Blanc thanks the Swiss National Science Foundation, 
and E. von Toerne thanks the Alexander von Humboldt Stiftung for support.
This work was supported by the National Science Foundation, the
U.S. Department of Energy, and the Natural Sciences and Engineering Research 
Council of Canada.

\end{document}